\documentclass[showpacs,twocolumn,preprintnumbers,amsmath,amssymb]{revtex4}

\usepackage{graphicx}
\usepackage{subfigure}
\usepackage{dcolumn}
\usepackage{bm}

\begin{document}

\title{Generic evaluation of the relaxation time to equilibrium}
\author{Takaaki Monnai}%
\email{monnai@suou.waseda.jp}%
\affiliation{$*$Department of Applied Physics, Osaka City University,
3-3-138 Sugimoto, Sumiyoshi-ku, Osaka 558-8585, Japan}
\begin{abstract}
We evaluate the relaxation time to equilibrium, and especially show that it is almost independent from the system size for macroscopic isolated quantum systems. It at most polynomially depends on the system size.
This estimation holds when the Hamiltonian is non-integrable, the initial deviation of the quantity of interest is of order its spectral norm, and the relaxation process is monotonic.
\end{abstract}    
\maketitle

Recently many attentions have been paid for the foundation of equilibrium statistical mechanics.  
For macroscopic quantum systems, the relaxation to equilibrium has been explained by probabilistic arguments based on the extreme high dimensionality of the Hilbert space stems from the noncommutativity\cite{Deutsch,Lebowitz,Sugita,Reimann,Popesucu,vonNeumann,Lebowitz2,Lebowitz4,Rigol,Reimann2,Popesucu2,Monnai}. Refs.\cite{Deutsch,Lebowitz,Sugita,Reimann,Popesucu} establish that majority of the states are regarded as equilibrium, whose properties are commonly shared. On the other hand, Refs.\cite{vonNeumann,Lebowitz2,Lebowitz4,Tasaki,Rigol,Reimann2,Popesucu2,Monnai} explain the relaxation process by also assuming either macroscopic properties of observables or nonintegrability of the Hamiltonian.
 
But the evaluation of the relaxation time remains as a significant unsolved problem.
It would strongly depend on models and quantities of interest, and thus is believed to be difficult to estimate.
Even very fundamental properties are in general unclear such as how short the relaxation time is compared with an extremely long recurrence time\cite{Peres}, or how strongly it depends on the quantities of interest given that the relaxation occurs.     

In this letter, we show a generic estimation of the relaxation time under the main assumptions i-iii).
The assumptions are as follows.
\begin{itemize}
\item[i)]The Hamiltonian is nonintegrable.
\item[ii)]For a quantity of interest $A$ which polynomially depends on the system size, the deviation $\delta A(0)$ of the initial expectation value from the long time average is less than a kind of spectral norm $\|A\|$.
\item[iii)]In the course of time evolution, the deviation of the expectation value $\delta A(t)$ monotonically approaches to zero.             
\end{itemize}
The assumption iii) implicitly assumes that the relaxation occurs.
Physically, the relaxation time is expected to be almost independent from the system size or volume for macroscopic systems. Indeed, macroscopic systems are divided into many pieces which are mutually interacting, and each portion would relax independently. From this macroscopic argument, the polynomial dependence at most would be trivial, however, here it is shown as a generic theorem.    

Let us consider a normalized state $|\psi(t)\rangle$ in a large Hilbert space ${\cal H}_{[E,E+\Delta E]}$ at an energy $E$.
We are concern with the relaxation time for a large but isolated system. 
Justifications for the use of isolated system to describe relaxation has been discussed for example in Ref.\cite{Reimann2}.  
From an initial nonequilibrium, it evolves by a time-independent Hamiltonian $H$ whose eigenenergies and eigenstates are $E_n$ and $|E_n\rangle$, respectively
\begin{equation}
|\psi(t)\rangle=\sum_{n=1}^d c_n e^{i\phi_n}e^{-\frac{i}{\hbar}E_n t}|E_n\rangle,
\end{equation}
where $c_n$ and $\phi_n$ are the amplitude and phase of eigenstate $|E_n\rangle$. 
$d={\rm dim}{\cal H}_{[E,E+\Delta E]}$ is the dimension of the Hilbert space, which exponentially depends on the system size $N$. The width $\Delta E$ is smaller than $E$, but much larger than neighboring energy distances of order $\frac{\Delta E}{d}$. Such a width is always necessary for the microcanonical approach. 
We consider a physical quantity $A$ whose maximum value $\|A\|={\rm Max}_{|\Phi\rangle\in {\cal H}_{[E,E+\Delta E]}}\frac{\langle\Phi|A|\Phi\rangle}{\langle \Phi|\Phi\rangle}$ polynomially depends on the system size $N$. For example, the sum of local finite quantities is proportional to $N$ and satisfy this condition, and these are important macroscopic quantities\cite{Sugita,Monnai}.   

The difference between the expectation value and long time average of a quantity of interest $A$ is calculated as
\begin{eqnarray}
&&\delta A(t) \nonumber \\
&=&|\langle\psi(t)|A|\psi(t)\rangle-\sum_{n=1}^d c_n^2\langle E_n|A|E_n\rangle| \nonumber \\ 
&=&|\sum_{n\neq m=1}^d c_n c_m e^{\frac{i}{\hbar}(E_n-E_m)t+i(\phi_n-\phi_m)+i\gamma_{nm}}|\frac{\|A\|}{\sqrt{d}} \nonumber \\
&=&|\sum_{n\neq m=1}^d e^{i(\phi_n-\phi_m)+i\gamma_{nm}+\frac{i}{\hbar}(E_n-E_m)t}|\frac{\|A\|}{d\sqrt{d}},\label{mean}
\end{eqnarray}
where we evaluated $c_n=O(\frac{1}{\sqrt{d}})$ from the normalization, and $|\langle E_n|A|E_m\rangle|=O(\frac{\|A\|}{\sqrt{d}})$ for nonintegrable systems\cite{Monnai}, which is correct including the standard deviation. Regarding the evaluation of $c_n$, it is also possible to say that we choose $c_n=\frac{1}{\sqrt{d}}$ and rigorously evaluate the order of the relaxation time. The broad distribution over the eigenstates is also reasonable for the realistic preparation of the initial states\cite{Reimann2}.  $\gamma_{n,m}=-\gamma_{m,n}$ is the phase of the off-diagonal element $\langle E_n|A|E_m\rangle$, which is in general a function of $(m,n)$.  Note that the diagonal contribution expresses the long time average 
\begin{eqnarray}
&&\sum_{n=1}^d c_n^2 \langle E_n|A|E_m\rangle \nonumber \\ 
&=&\lim_{T\rightarrow\infty}\frac{1}{T}\int_0^T \sum_{n,m=1}^d c_n c_m e^{i(\phi_n-\phi_m)+\frac{i(E_n-E_m)}{\hbar}t}\langle E_n|A|E_m\rangle. \nonumber \\
&&
\end{eqnarray}
Also note that our analysis only requires that there is a typical small amplitude of off-diagonal elements.
And we shall skip the quantitative evaluation of the off-diagonal elements since it is detailed in Ref.\cite{Monnai} with numerical evidences, but its smallness is explained as below.
Intuitively, the off-diagonal elements are extremely small compared to the diagonal ones, since they are related to the transition amplitudes between macroscopically different states $|E_n\rangle$ and $|E_m\rangle$ by a perturbation $A$, which is an almost impossible process. Note that the argument here holds also for $E_n$ very near to $E_m$, since the shape of the corresponding eigenvectors are different.    
 
The initial state shows deviation from equilibrium value, and we assume
\begin{equation}
|\sum_{n\neq m=1}^d e^{i(\phi_n-\phi_m)+i\gamma_{nm}}|= d^{\alpha},\label{phase}
\end{equation}
where the important parameter $\alpha$ is determined later.
Note that if the phases $\{\phi_n-\phi_m+\gamma_{nm}\}$ are uncorrelated, the sum over the random phases corresponds to $d(d-1)$ steps of a unbiased random walk, and thus $\alpha=1$.
To clarify this point, let us consider a uniformly random quantity $\xi_{nm}$ and estimate 
$|\sum_{n,m=1}^d e^{i\xi_{nm}}|$.
For a fixed $m=m_0$, ${\rm Re} e^{i\xi_{nm_0}}$ is also unbiased random variable taking positive and negative values with an equal probability. The summation over $n$ is therefore of order $\sqrt{d}$ and can be both positive and negative. Then the sum over $m$ is actually regarded as a random walk with an almost common step $\sqrt{d}$, which yields total displacement $O(d)$, i.e. $\alpha=1$. 

Substituting the ansatz Eq.(\ref{phase}) to the evaluation of expectation value Eq.(\ref{mean}), 
we have 
\begin{equation}
|\langle\psi(0)|A|\psi(0)\rangle-\sum_{n=1}^d c_n^2\langle E_n|A|E_n\rangle|=d^{\alpha-\frac{3}{2}}\|A\|.
\end{equation}
We assume that the initial deviation $\delta A(0)$ is of the same order as $\|A\|$. 
Therefore the exponent satisfies $\alpha=\frac{3}{2}-\epsilon$ with a very small positive quantity $\epsilon=+0$.
$\epsilon$ accounts for the case that the deviation is slightly smaller than $\|A\|$.
We use this notation for $\epsilon$, since $d=O(e^N)$ is extremely large number and $\epsilon$ can be smaller than any positive number for sufficiently large $N$.   

For $\alpha=\frac{3}{2}-\epsilon$, there is a strong correlation among the phases $\{\phi_m\}$ compared with uncorrelated case $\alpha=1$. In this paragraph, we estimate how long the phase  correlation is maintained.
Phases $S=\{\phi|\phi=\phi_n-\phi_m+\gamma_{nm}, 0\leq m,n\leq d\}$ are decomposed to $S_1$ whose elements are concentrated around a specific value $0\leq\phi\leq 2\pi$  and a completely uncorrelated set $S_2=S-S_1$ whose elements randomly distributed over $[0,2\pi]$.
Strictly speaking, there can be several $\phi$, around which the correlated components of $S_1$ distribute. But for simplicity, we assume single valued $\phi$, which does not affect our analysis.   
In the course of time evolution, $S_1$ is well-defined when the additional phases satisfy $\frac{|E_n-E_m|}{\hbar}t\ll \phi$ for most $n$ and $m$. This condition is satisfied for $t\leq\frac{\hbar}{\Delta E}\phi$, the decorrelation time scale for $n$ and $m$ with $|E_n-E_m|$near $\Delta E$. Other additional phases with much smaller $|E_n-E_m|\leq\Delta E$ does not affect the correlation of $S_1$ within the duration. We also required that $S_2$ remains uncorrelated.

Let us calculate the relaxation time $T_r$.
The time derivative of the expectation value is calculated as
\begin{eqnarray}
&&|\frac{d}{dt}\langle \psi(t)|A|\psi(t)\rangle| \nonumber \\ 
&=&2\frac{E}{\hbar}|\sum_{n\neq m=1}^d \sin(\phi_n-\phi_m+\gamma_{nm}+\frac{E_n-E_m}{\hbar} t)|^2\frac{\|A\|}{d\sqrt{d}} \nonumber \\
&=&O(\frac{E\|A\|}{\hbar}d^{\alpha-\frac{3}{2}}) \label{derivative}
\end{eqnarray}
for $t\leq\frac{\hbar}{\Delta E}\phi$.
This provides the rate of relaxation.
Now we assume that the expectation value monotonically approaches to the equilibrium.
The relaxation time required for this change of the expectation value is evaluated as $T_r\frac{E\|A\|}{\hbar}d^{\alpha-\frac{3}{2}}=\|A\|$ from Eq.(\ref{derivative}), and thus 
\begin{equation}
T_r=\frac{\hbar}{E}d^{\frac{3}{2}-\alpha}.\label{relaxation}
\end{equation}

The relaxation time is therefore evaluated as
\begin{equation}
T_r=\frac{\hbar}{E}d^{\epsilon}.\label{relaxation}
\end{equation}
Therefore from the definition of an extremely small parameter $\epsilon$, the dependence on $d$ of the relaxation time is weaker than any polynomials of $d$. The logarithmic dependence on $d$ is possible, and $T_r$ at most polynomially depends on the system size $N$. This includes the realistic case where the relaxation time $T_r$ is independent from system size. 
More generally, $T_r=\frac{\hbar}{E}N^a$ with $a\geq 0$ if we choose $\epsilon=\frac{a\log\log d}{\log d}$ which can be smaller than any positive quantity as $N\rightarrow\infty$.     

In conclusion, we give the first step of the studies for generic properties of the relaxation time. We evaluated the relaxation time and showed that it at most polynomially depends on the system size for generic systems satisfying assumptions i-iii). The evaluation guarantees a reasonable relaxation time without going into the detail of a specific model. 

The author is grateful to Professor A.Sugita for valuable discussions. 
This work is financially supported by JSPS research program under the Grand 22$\cdot$7744. 

\end{document}